\documentclass[aps, amssymb, amsmath, superscriptaddress, prl,
twocolumn] {revtex4-2}

\usepackage{graphicx}
\usepackage{color}
\usepackage{amsmath}
\usepackage{enumitem}
\usepackage{amssymb}
\usepackage{hyperref}

\newcommand{\be}{\begin{equation}}
	\newcommand{\ee}{\end{equation}}
\newcommand{\bea}{\begin{eqnarray}}
	\newcommand{\eea}{\end{eqnarray}}
\newcommand{\p}{\partial}

\newcommand{\lb}{\left[}
\newcommand{\rb}{\right]}
\newcommand{\lp}{\left(}
\newcommand{\rp}{\right)}

\renewcommand{\Re}{{\rm \, Re\,}}

\renewcommand{\vec}[1]{{\boldsymbol #1}}

\begin{document}

	\title{Spontaneous Running Waves and Self-Oscillatory Transport in Dirac Fluids}
	
	\author{
Prayoga Liong}\thanks{Equal contributions}\affiliation{Department of Physics,  Boston University, Boston MA 02215, USA} 

\author{Aliaksandr Melnichenka}\thanks{Equal contributions}\affiliation{Department of Physics, Berea College, Berea, KY 40404, USA}

\author{Anton Bukhtatyi}\affiliation{Department of Physics, V. N. Karazin Kharkiv National University, Kharkiv, Ukraine}
\author{Albert Bilous}\affiliation{Department of Physics,  Taras Shevchenko National University, Kyiv, Ukraine} 

\author{Leonid Levitov}\affiliation{Department of Physics, Massachusetts Institute of Technology, Cambridge MA 02139, USA} 

\begin{abstract}
We predict 
hydrodynamic Turing 
instability of current-carrying 
Dirac electron fluids that drives spontaneous self-oscillatory transport. The instability arises near charge neutrality, where carrier kinetics make current dissipation strongly density dependent. Above a critical drift velocity, a uniform electronic flow becomes unstable and undergoes a dynamical transition to a state with coupled spatial modulation and temporal oscillations—an electronic analogue of Kapitsa roll waves in viscous films. The transition exhibits two clear signatures: a nonanalytic, second-order–like onset in the time-averaged current and narrow-band electromagnetic emission at a tunable washboard frequency $f=u/\lambda$. Although reminiscent of sliding charge-density waves, the mechanism is intrinsic and disorder independent. Owing to the small effective mass of Dirac carriers, hydrodynamic time scales translate into emission frequencies in the tens to hundreds of gigahertz range, establishing Dirac materials as a platform for high-frequency self-oscillatory electron hydrodynamics.
\end{abstract}
\date{\today}

\maketitle

Pattern formation is a hallmark of driven many-body systems, revealing how collective behavior can emerge even without external structuring. A paradigmatic example is the Turing instability~\cite{Turing1952,Murray2002,Prigogine1977}. Despite its universality, this mechanism has been explored almost exclusively in classical settings. Here we predict a quantum analogue: a modulational (Turing-type) instability in quantum materials driven out of equilibrium by a DC current. The key idea is that an applied current selectively softens the system at particular wave numbers, triggering an instability toward spontaneous spatial modulation. We find that current-driven Turing effects naturally arise in quantum dissipative systems, such as Dirac and Weyl metals in which electron-hole scattering enhances resistivity near charge neutrality while remaining weak away from it~\cite{Fritz2008,Muller2008,Kashuba2008,Gallaher2019,Ku2019,Lucas2017,Morpurgo2017,Lucas2018}.

Beyond fundamental interest, determining when a current-carrying state becomes unstable to self-oscillation may also clarify a longstanding practical question: Can a DC current generate coherent high-frequency radiation? Proposed mechanisms have ranged from Bloch oscillations \cite{Bloch1929,Kittel1963,Feldmann1992,Leo1998} to plasmonic instabilities \cite{Dyakonov1993,Dyakonov1995}, and while experimental signatures exist \cite{Kastrup1995,Roskos1992,Knap2002,Knap2006,Boubanga2010}, achieving robust and tunable self-oscillation has remained difficult. Our results identify a new route to current-driven pattern formation and self-oscillation--- arising from the interplay of dissipation, band geometry, and hydrodynamic effects in Dirac materials --- and highlight fresh opportunities for exploring quantum systems with complex band geometry or strong correlations.

\begin{figure}[tb]
    \centering
    \includegraphics[width=0.95\linewidth]{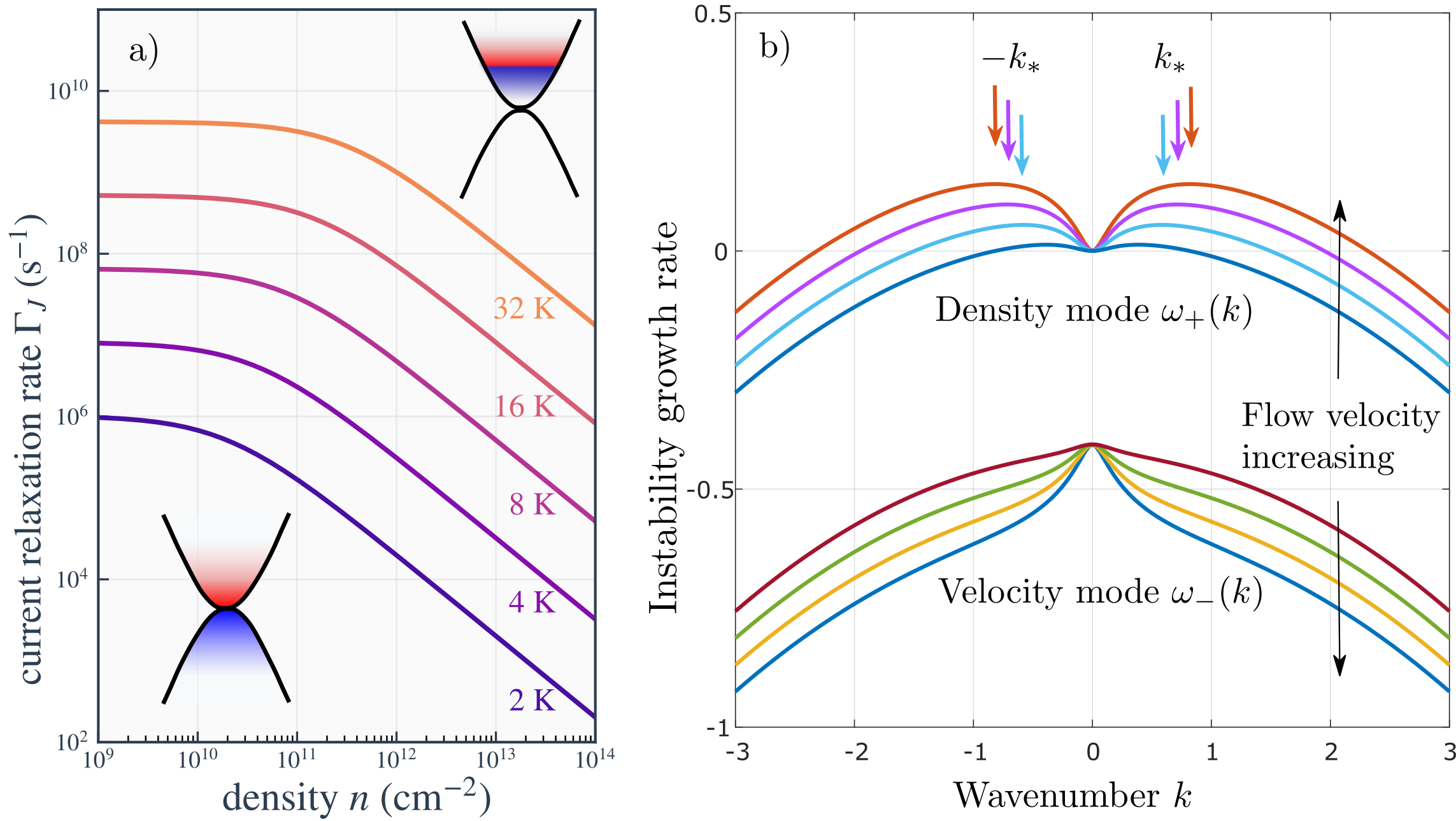} 
    \caption{(a) Current dissipation 
    in graphene bilayer (BLG)  band vs. electron density. Dissipation is high in the Dirac plasma near charge neutrality point (CNP) but drops rapidly upon doping away from CNP. 
    Density-dependent dissipation drives the Turing–Kapitsa instability. Insets illustrate carrier population in BLG band for densities at CNP and away from it, respectively. 
    (b) Instability growth rate ${\rm Im}\,\omega_\pm(k)$ for the modes $\omega_+(k)$ and $\omega_-(k)$, Eq.~\eqref{eq:modes1}, representing density and velocity perturbations. The density mode satisfies $\omega_+(0)=0$ by particle-number conservation, becoming unstable above a critical flow velocity (Eq.\eqref{eq:u_critical}); the growth rate ${\rm Im}\,\omega_+(k)>0$ peaks at the wave numbers $\pm k_*$ marked by arrows. The velocity mode remains damped, ${\rm Im}\,\omega_-(k)<0$. Small initial perturbations grow into a periodic modulation with wavelength $\lambda\sim 2\pi/k_*$ propagating downstream, as shown in Fig.~\ref{fig3}.}
\vspace{-4mm}
    \label{fig2}
\end{figure}

The instability mechanism proposed here is reminiscent of 
the Kapitsa instability in viscous films flowing down an incline \cite{Craster2009, Kapitza1948,Andreev1964,Balmforth2004,Chimetta,video}. The emergence of periodic modulations in steady laminar flows of open-surface films is a direct consequence of the balance between gravity and viscous drag: $
\rho g \sin\theta = \eta \nabla^2 \vec{v}(\vec{r}) $, 
where $\theta$ is the incline angle, $g$ gravity, and $\eta$ viscosity. A steady state defines a scaling between flow velocity and film thickness, $u \propto w^2 g/\eta$. Since 
$u$ grows rapidly with $w$, regions with slightly larger film thickness $w(x)$ will flow faster, thereby amplifying perturbations and giving rise to a self-sustained running wave,
\be
w(x) = \bar{w} + \delta w \cos k(x - ut).
\ee
In fluid films, the wavelength $\lambda = 2\pi/k$ and modulation amplitude are set by the competition between viscous and capillary forces \cite{Kapitza1948}. The emerging wavenumber $k$ coincides with $k_*$—the wavevector at which the instability growth rate is maximal—mirroring the essence of Turing’s pattern-formation mechanism \cite{Andreev1964}.

As we will see, an electronic analogue of this instability can cause a steady (DC) electron flow to spontaneously develop self-sustained charge and current oscillations. The correspondence between an atomically thin 2D conductor with carrier density $ n(x,y) $ and a classical viscous film with spatially varying thickness $ w(x,y) $—measured between the substrate and its free surface—is established by identifying $ n(x,y) $ with $ w(x,y) $. Beyond this “hyperspace mapping” $w(x,y) \to (x,y,n)$, realizing the Turing–Kapitsa instability requires a dissipation mechanism whose rate, being strong, varies rapidly with $ n $—the electronic analogue of a drag coefficient in viscous films that decreases sharply with increasing film thickness.

\begin{figure}[tb]
    \centering
    \includegraphics[width=0.95\linewidth]{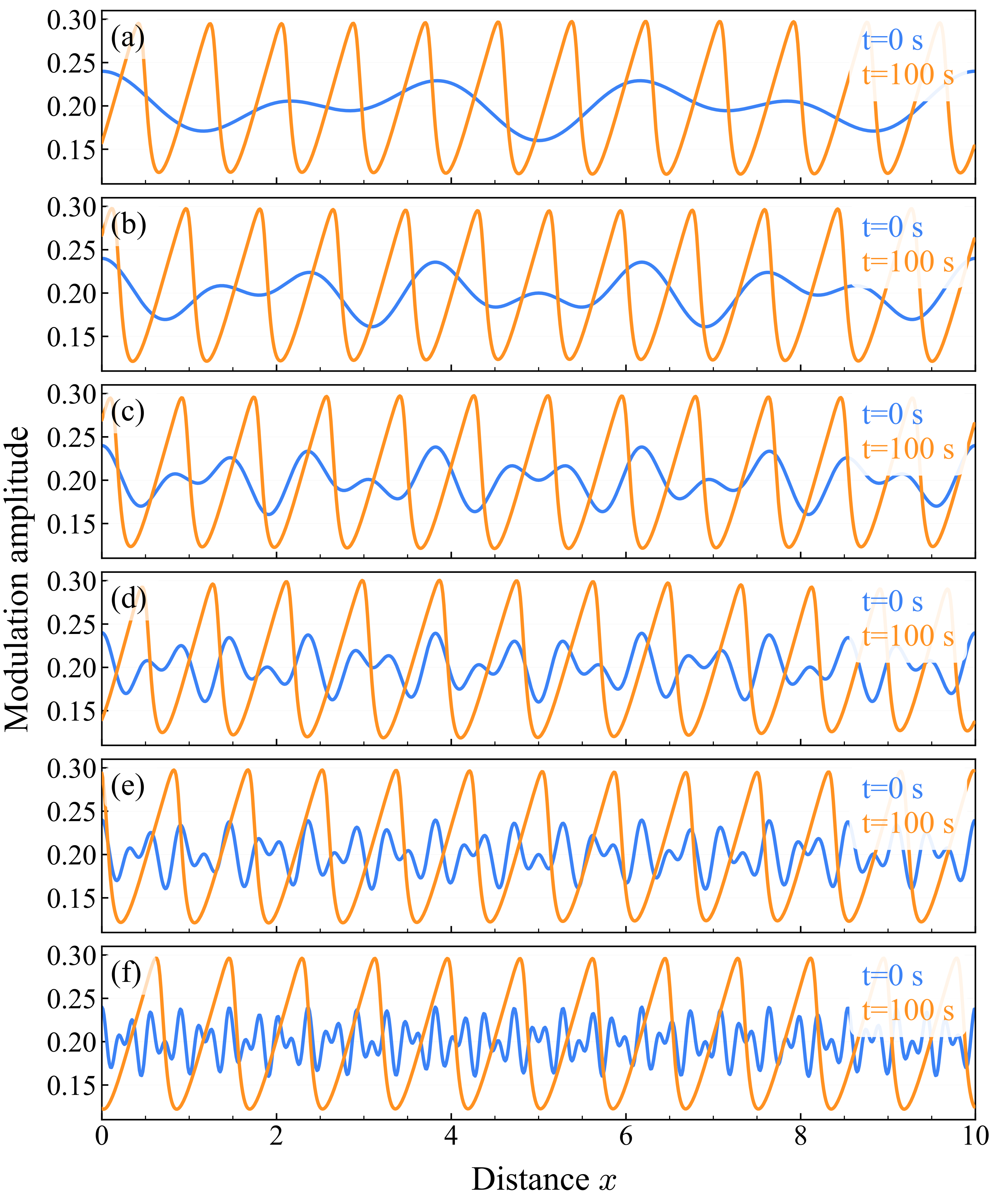} 
    \caption{Wavelength selection at instability. Shown is the long-time modulation (orange) evolved from different initial perturbations (blue):
$\delta n(x)=\sum_m a_m\cos k_m x+b_m\sin k_m x$, $\delta p(x)=\sum_m a'_m\cos k_m x+b'_m\sin k_m x$, with $k_m=\frac{2\pi}{L}\times 3,5,8,13,21,55$. 
Irrespective of initial conditions, the system converges to the same spatial pattern, with wavelength set by the wavenumber $k$ of maximal growth (marked by arrows in Fig.~\ref{fig2}), consistent with Turing’s maximum-growth rule.
    }
  \vspace{-7mm}
    \label{fig3}
\end{figure}

Several effects reported in the literature can cause dissipation to decrease as carrier concentration increases. Here we focus on quasiparticle scattering at charge neutrality (CNP) \cite{Fritz2008,Muller2008,Kashuba2008,Gallaher2019,Ku2019,Lucas2017,Morpurgo2017,Lucas2018}. This dissipation channel is strong at CNP, where the Dirac plasma is e--h balanced, but it weakens rapidly once the carrier density is tuned away from neutrality. The resulting density-dependent dissipation plays a role similar to the width-dependent viscous drag in thin-film hydrodynamics: when the drift velocity exceeds a threshold, the flow becomes unstable and develops downstream-propagating waves. 

Our analysis predicts a critical drift velocity: 
\begin{equation}
u_c=\frac{u_0}{R}, \qquad
R=\frac{n}{\gamma} \left|\frac{d\gamma}{d n}\right|,
\qquad
u_0=\sqrt{\frac{Un}{m}},
\label{eq:u_critical}
\end{equation}
where 
the dimensionless quantity $R$ measures how sensitively the dissipation rate $\gamma$ responds to changes in carrier density, while $u_0$ denotes the 2D plasmon velocity at density $n$. Because the dissipative regime of interest lies close to the charge neutrality point (CNP), the relevant carrier densities are much smaller—i.e., much closer to neutrality—than those typically used in studies of ballistic 2D plasmon propagation \cite{Fei2012,Woessner2015,Zhao2023_energywaves}. As shown in Fig.~\ref{fig2}(a), the strong temperature dependence of $\gamma$ produces a pronounced density dependence of $R$ near the CNP. As a result, the critical velocity $u_c$ varies strongly with both temperature and carrier density \cite{Supplemental Information}.

\begin{figure}[tb]
    \centering
    \includegraphics[width=0.97\linewidth]{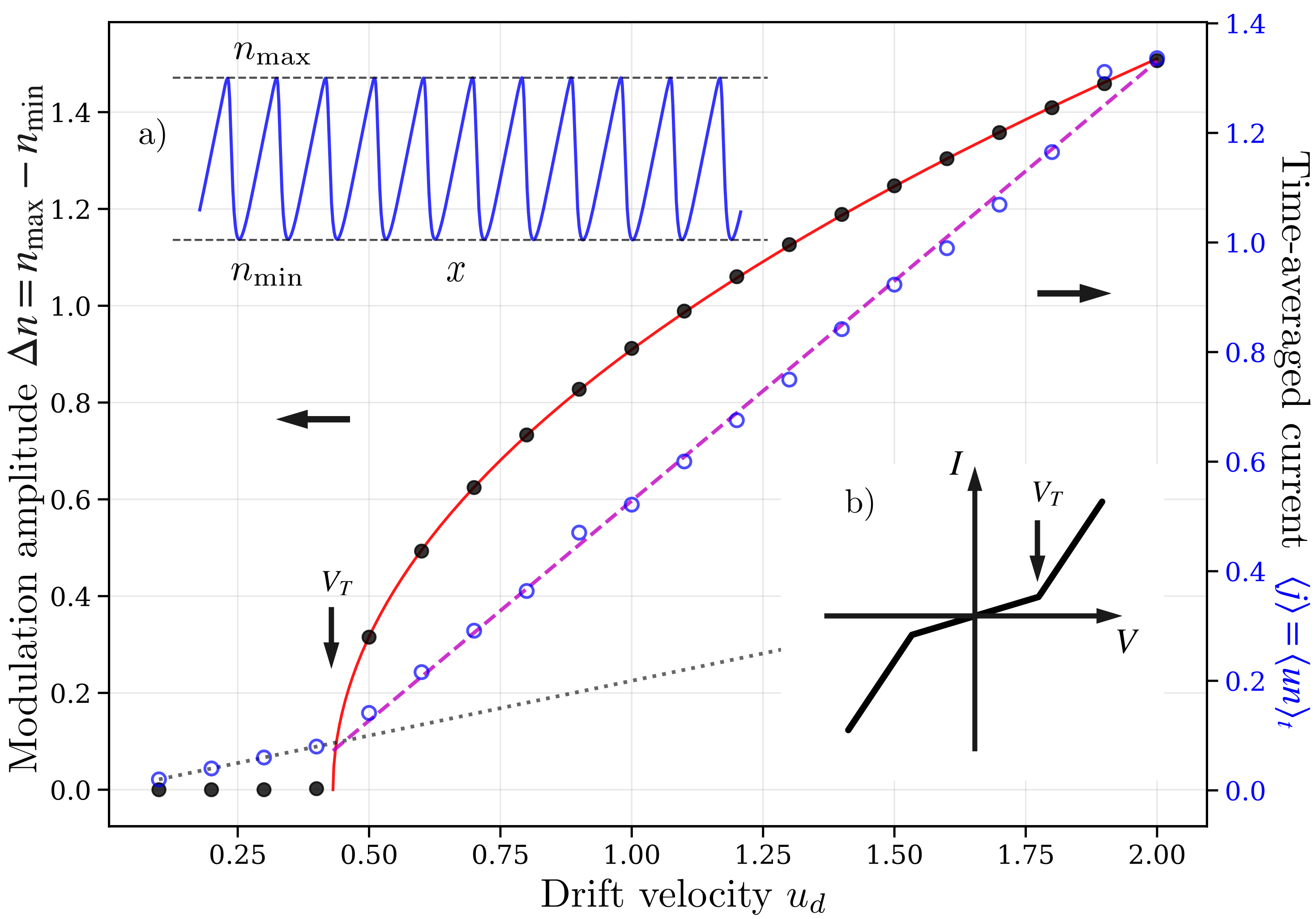} 
    \caption{Signatures and observables for current-driven electronic Turing–Kapitsa instability. 
    Plotted are the running waves modulation amplitude and time-averaged current near the instability onset, obtained numerically for a graphene bilayer model (see Eqs.~\eqref{eq:transport_eqs_p}, \eqref{eq:transport_eqs_n} and accompanying text). The red line shows the theoretical square-root behavior $\Delta n\propto (u_d-u_d^*)^{1/2}$ expected on the general symmetry grounds \cite{LL Fluid Mechanics}.  
    The sharp slope change in the time-averaged (DC) current–field dependence marks the instability threshold and the added current from the running waves. Insets (a) and (b) illustrate a typical modulation and the current–field relation.
    }
\vspace{-4mm}
    \label{fig1}
\end{figure}

For $u>u_c$, the system develops a self-sustained modulation that follows Turing’s wavelength-selection rule. Close to threshold, the modulation wavelength is $\lambda = 2\pi/k_*$, where $k_*$ is the most unstable mode, at which the instability growth rate is maximal (see Figs.~\ref{fig2}(b) and~\ref{fig3}). The transition is continuous and second-order (Fig.~\ref{fig1}), and the modulation wavelength remains finite at the threshold, as in Brazovskii and Swift-Hohenberg theories of convective instabilities\cite{Brazovskii1975,Swift_Hohenberg1977}, without showing any divergence. For realistic parameters, 
$\lambda$ spans a wide range, from submicron to hundreds of microns \cite{Supplemental Information}.



Observable signatures of this collective dynamics resemble those studied for other running wave phenomena, such as sliding charge-density and spin-density waves (CDW and SDW) \cite{LeeFukuyama, LeeRice, GrunerRMP}. One is a narrow-band emission at the ``washboard'' frequency 
$f=u/\lambda$, where $u$ is the flow velocity tunable by the applied DC current.  Another is a characteristic nonlinearity 
in the current–field dependence at the instability threshold, shown in Fig.\ref{fig1}.
The increase in the slope of the current vs. $u$ dependence
reflects the contribution of running waves to nonlinear conductivity, resembling the Lee-Fukuyama-Rice conductivity of sliding CDW\cite{LeeFukuyama, LeeRice, GrunerRMP}. 

We stress that both the AC emission by running waves at a frequency $f=u/\lambda$ and the DC current nonlinearity are intrinsic effects, present already in a homogeneous system. This is in contrast to CDW sliding, where narrow-band emission and current nonlinearity are governed by the effects due to pinning-depinning by disorder. 

To explore the emergence of running waves under Turing-Kapitsa instability driven by a DC current, 
we consider a simple but realistic transport model. 
A natural  framework to analyze collective dynamics of electrons 
is provided by hydrodynamic equations for particle number and momentum density $n(x,t)$ and $\vec p(x,t)$. Microscopically, these quantities are 
\be \label{eq:n_p_definition}
 n( x,t)=\sum_{\vec p'} f(\vec p',x,t),\quad
\vec p(x,t) =\sum_{\vec p'} \vec p' f(\vec p',x,t)
,
\ee
where 
$f(\vec p,x,t)$ is particle phase-space distribution 
 summed over spins and valleys that evolves in time under the kinetic equation \cite{LL10,Tong}
\be\label{eq:kinetic_eqn}
(\p_t+\vec v\cdot \p_{\vec x}+\vec F\cdot \p_{\vec p}) f(\vec p,\xi)=I[f]
.
\ee
Here, $\vec v=\p_{\vec p}\epsilon(p)$ is the velocity, $\vec F=e\vec E$ the force, and $I[f]$ the collision integral describing electron-electron and electron-hole scattering, and interactions with 
disorder and phonons; $\xi=(x,t)$ denotes space and time variables.  

We are interested in transport in a 
Dirac band of a graphene bilayer, $ 
\varepsilon_{\pm}(\vec k)
    = \pm \tfrac12\!\left(\sqrt{\gamma_1^2 + 4(\hbar v k)^2} - \gamma_1\right)$,
where $v\approx 10^6\,$m/s and $\gamma_1=0.39\,$eV.
For simplicity, we consider a one-species model representing 
one component of the bilayer bands---either electron or hole.  
The results below are largely insensitive to the exact form of $\epsilon(p)$ or to microscopic details of how momentum dissipation varies with carrier density. We model this by a dissipation rate that peaks in the region of width $w\sim k_B T$ and decreases with detuning from the CNP\cite{Morpurgo2017,Lucas2018}:
\be\label{eq:Gamma_n}
\gamma(n) = \gamma_0 e^{-|n|/w}, \quad
w = \frac{2m }{\hbar^2}k_B T 
, \quad
m\approx 0.04 m_{\rm e}, 
\ee
where 
$w$ describes the CNP width due to intrinsic e-h scattering in pristine graphene in a parabolic band approximation. 
In this model, the electron temperature is replaced by an effective time-averaged value, neglecting its dynamics. Although heating is important, it occurs on time scales much longer than those of oscillatory electron motion. Moreover, recent experiments\cite{Berdyugin2022,Nowakowski2024,Geurs2025} show that high currents can flow through moiré graphene at drift velocities approaching the band velocity without significant overheating. Efficient cooling via the hBN substrate enables such high-current operation without damaging the graphene or broadening the Dirac band at CNP. 

When applied to the particle number and momentum densities, $n$ and $p$, Eq.\eqref{eq:kinetic_eqn} gives conservation laws:
\begin{align}
\lp \gamma(n)+\p_t\rp &\vec p(\xi)+\p_{\vec x}\Pi_0(\xi)=en  \lp \vec E-\p_{\vec x}\Phi (\xi)\rp
\label{eq:transport_eqs1} \\ \label{eq:transport_eqs0} 
\p_t & n(\xi)+\p_{\vec x} [n(\xi)\vec v(\xi)] =0
.
\end{align}
Here the first equation is the continuum version of the second Newton's law $dp/dt=F$, the second equation is the continuity equation for $n$. 
In a parabolic band, momentum density is related to particle current density $\vec j=\sum_{p'} \vec v(p') f(\vec p')\equiv n\vec v$ 
as $m\vec j=\vec p$. 


The quantity $\Pi_0$ is 
the Fermi pressure of a flowing Fermi sea described by Doppler-shifted Fermi distribution 
$
f(\vec p)=\frac1{e^{\beta(\epsilon(p)-\vec u\cdot\vec p-\mu)}+1}
$. 
For electron gas of density $n$ flowing with the velocity $u$, for parabolic band, $\Pi_0$ is a sum 
of a static contribution and a kinetic energy term 
\begin{align}\label{eq:FS_energy}
&\Pi_0 
=\sum_{p'} p_x' v_x(p')f(\vec p')=\frac{n^2}{2\nu}
+\frac{p^2}{nm}
,\quad p=mnu
,
\end{align} 
where $\nu=n/\epsilon_F=gm/2\pi\hbar^2$ is the density of states ($g$ is spin and valley degeneracy). 
%
The simple form of the result  in Eq.\eqref{eq:FS_energy}, which is a sum of the pressure of a Fermi sea at rest $\Pi_0(n)=\sum_{p'}\epsilon(p')f_{u=0}(p')=n^2/2\nu$ 
and the Bernoulli pressure 
$p^2/nm$ due to the flow, is a property of a parabolic band arising from its Galilean symmetry.


The transport equations 
\eqref{eq:transport_eqs1} and \eqref{eq:transport_eqs0} 
capture the essential physics underlying
the current-driven Turing instability. 
A more realistic description, however, must explicitly include the interaction and viscous effects incorporated in 
the hydrodynamic stress density $\Pi$.

Below, for conciseness, we use a constant interaction model 
deferring the discussion of 
long-range interaction till later.
In this case, the last term in 
\eqref{eq:transport_eqs1} takes the form 
\be\label{eq:Phi(x)}
\Phi(\xi)=U_0 [n(\xi)-\bar n]
,
\ee
where constant $\bar n$ represents an 
offset charge on a gate. 
 

It is convenient to combine the terms $\p_{\vec x}\Pi_0$ and $\p_{\vec x}\Phi$ in Eq. \eqref{eq:transport_eqs1} into a net  
Fermi pressure term  $\p_{\vec x}\Pi$ that accounts for both electron–electron interactions and the flow, \cite{footnote_Pi_Phi}
\be\label{eq:Pi_total}
\Pi=\frac12 U \lp n(\xi)\rp^2+\frac{(p(\xi))^2}{mn(\xi)}
,\quad
U=U_0+\frac1{\nu}
,
\ee
where $n$ and $p$ are particle and momentum densities. 
After these rearrangements, 
transport equations read
\begin{align}
& \lp \gamma(n)+\p_t\rp \vec p(\xi)+\p_{\vec x}\Pi(\xi)=en  \vec E 
\label{eq:transport_eqs_p} 
\\
& \p_t n(\xi)+\frac1{m}\p_{\vec x} \vec p(\xi)=0 
,
\label{eq:transport_eqs_n}
\end{align}
where in the continuity equation we replaced $nv$ with $p/m$. 
A steady uniform current-carrying state obeys 
\be
\gamma \vec p=en\vec E, 
\ee
yielding a 
standard result for the flow velocity, 
$
\gamma u =\frac{e}{m}  E
$.

Next, we show that 
Eqs. \eqref{eq:transport_eqs_p} and \eqref{eq:transport_eqs_n} 
predict instability of a current-carrying state. In performing numerics, it is beneficial to regularize the dynamics at large wavenumbers by introducing diffusion-like terms in the $p$ and $n$ equations. This can be done by replacing 
\be
\p_t p\to (\p_t-D_p\p_x^2) p,\quad
\p_t n\to (\p_t-D_n\p_x^2) n
\ee
in Eqs.\eqref{eq:transport_eqs_p} and \eqref{eq:transport_eqs_n}. 
This change does not affect the behavior 
at long wavelengths, provided $D_n$ and $D_p$ are small, however it makes the dynamics well behaved at short wavelengths, allowing to perform numerical simulations and explore the behavior at long times where the nonlinear behavior triggered by Turing instability emerges---see Figs.~\ref{fig2} and \ref{fig3}, and discussion in \cite{Supplemental Information}.    

Next, we examine the conditions under which the transport equations develop an instability. 
Perturbing the steady state as $n=\bar n+\delta n$, $p=\bar p+\delta p$, and 
linearizing Eqs.\eqref{eq:transport_eqs_p} and \eqref{eq:transport_eqs_n} in $n$ and $p$, gives
\begin{align} 
&\lp 
\gamma_n \bar p-eE\rp  \delta n 
+
\Pi_n \p_x \delta n 
+ \gamma\delta p+(\p_t  - D_p\p_x^2) \delta p
\nonumber
\\
\label{eq:eqs_linearized_2}
& +
\Pi_p \p_x\delta p  = 0
,\quad 
(\p_t -D_n\p_x^2 ) \delta n+\frac1{m}\p_x\delta p =0
,
\end{align}
where the quantities with  subscripts $n$ and $p$ denote partial derivatives: $\gamma_n=\p \gamma/\p n$, $\Pi_n=\p \Pi/\p n$ and $\Pi_p=\p \Pi/\p p$. 
%
%
%
For harmonic 
perturbations $\delta n(\xi)=\delta n_{\omega,k} e^{ikx-i\omega t}$, $\delta p(\xi)=\delta p_{\omega,k} e^{ikx-i\omega t}$ this problem yields a pair of coupled linear equations 
\begin{align}
(\Lambda+ik\Pi_n)\delta n_{\omega,k}&+(\gamma-i\omega+D_pk^2 +ik\Pi_p)\delta p_{\omega,k} =0&
\nonumber 
\\
 (-i\omega+D_nk^2)&\delta n_{\omega,k} +i\frac{k}{m}\delta p_{\omega,k} =0
\end{align}
where we used a shorthand notation 
$\Lambda=
\lp \frac{\p \gamma}{\p n}-\frac{\gamma}{n}\rp p$. 
Instability occurs when 
the linearized problem admits a complex-frequency solution growing with time, 
\be
\delta n_k(t),\, \delta p_k(t)\sim e^{-i\omega t},\quad 
{\rm Im}\, \omega_+(k)>0.
\ee
The characteristic equation for this problem, in matrix form, reads 
\be
\lp \begin{array}{cc}
\Lambda 
+ik \Pi_n & \gamma-i\omega+D_pk^2 +ik \Pi_p
\\
-i\omega +D_nk^2 & i \frac{k}{m}
\end{array}\rp
\lp
\begin{array}{c}
\delta n_{\omega,k}
\\
\delta p_{\omega,k}
\end{array}
\rp=0
.
\ee
This yields two normal modes with frequencies 
\be\label{eq:modes1}
\omega_{\pm}(k)=\frac{-i \tilde\gamma+k\Pi_p 
\pm i\sqrt{\Delta}}{2} -iD_nk^2
,
\ee
where $\tilde\gamma=\gamma+(D_p-D_n)k^2$ and
\be\label{eq:modes2}
\Delta=\lp \tilde \gamma+ik\Pi_p\rp^2+ \frac{4ik}{m} \Lambda-\frac{4k^2}{m}\Pi_n
.
\ee
This general result is applicable to systems with both low and high dissipation, 
and both in the absence and in the presence of the flow. 
At $k=0$ the frequency $\omega_+$ vanishes, whereas the frequency $\omega_-$ takes a negative-imaginary value, corresponding to the conserved density and relaxing velocity modes, respectively.

\begin{figure}[tb]
    \centering
    \includegraphics[width=0.8\linewidth]{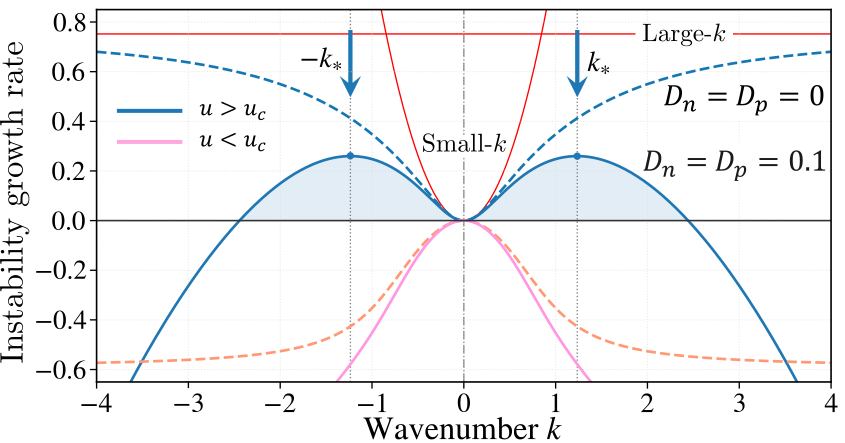} 
    \caption{Modulation wavelength $\lambda = 2\pi/k_*$, where $\pm k_*$ (blue arrows) maximizes the instability growth rate $\mathrm{Im}\,\omega_+(k)$, is governed by the behavior of $\mathrm{Im}\,\omega_+(k)$ at small and large $k$. For small $k$, the growth rate (red parabola) behaves as $\alpha k^2$, with $\alpha>0$ for $u>u_c$ and $\alpha<0$ for $u<u_c$. For large $k$, it saturates to a $k$-independent value (horizontal red line), again positive for $u>u_c$ and negative for $u<u_c$. Matching these small- and large-$k$ asymptotics yields Eq.~\eqref{eq:zeta_large-k}, 
    and hence Eq.~\eqref{eq:k*_result}. The instability growth rate is shown for $u=0.55>u_c$ and $u=0.2<u_c$ (blue and pink curves); solid and dashed lines correspond to $D_{n,p} = 0.1$ and $D_{n,p}=0$.
    }
\vspace{-4mm}
    \label{fig4}
\end{figure}


Prior to analyzing 
the dissipative regime $\gamma \gg \omega$—the main focus of this work—we 
verify that, in the weak-dissipation limit, our results reproduce the known plasma mode dispersions, both with and without steady background flow.
Setting $\gamma$, $\Lambda$, $D_n$ and $D_p$ to zero gives
\be
\omega_{\pm}(k)=\frac12 k\Pi_p 
\pm \sqrt{\frac14 k^2\Pi_p^2+
\frac{1}{m} k^2\Pi_n}
,
\ee
where we moved $i$ under the square root. Focusing on a parabolic band and plugging $\Pi_p=2p/nm$ and $\Pi_n=Un-p^2/n^2m$ [with $U$ the sum of the ee interaction and Fermi sea pressure contributions, as given in Eq.\eqref{eq:Pi_total}] the $p$ dependence under the square root cancels, giving 
\be\label{eq:Doppler_shift}
\omega_{\pm}(k)=ku 
\pm k u_0 
,\quad u_0=\sqrt{Un/m}.
\ee
Here $u_0$ is plasmon velocity at density $n$ and $u$ is the flow velocity, $p = m n u$. This gives the Doppler-shifted undamped plasma dispersion $(\omega - u k)^2 = k^2 u_0^2$, as expected from Galilean symmetry for a parabolic band.



Next, we estimate maximally unstable wavelength and instability threshold for the dissipative regime, where the relaxation rate 
$\gamma$ greatly exceeds scales such as $\omega p $ and $k\Pi $. 
For a parabolic band model 
the general expression for 
$\omega_\pm$ given in Eqs. \eqref{eq:modes1} and \eqref{eq:modes2} simplifies considerably. Substituting the derivatives $\Pi_n$ and $\Pi_p$ in Eq. \eqref{eq:modes2} yields 
\be
\Delta=\gamma^2+ik\frac{4p}{m}\frac{d\gamma}{dn}-4k^2 u_0^2 
\ee
%
%
An analytic estimate of the most unstable wavenumber $k_\ast$ can be obtained as illustrated in Fig.~\ref{fig4}. We compare the instability growth rate ${\rm Im}\,\omega_+(k)$ in the small- and large-$k$ limits and match these asymptotic behaviors in an intermediate (“borderline”) region. For clarity, we focus on the case of small $D_n$ and $D_p$. In the large- and small-$k$ limits, the growth rate behaves as
%
\be\label{eq:zeta_large-k}
{\rm Im}\,\omega_+(k)
=
\begin{cases}
-\frac12\gamma+\frac{\left|  p\frac{d\gamma}{dn} \right|}{2 mu_0}  
, & k\gg k_*,
\\
k^2\frac{\lb p^2 \lp \frac{d\gamma}{dn}\rp^2 
-m^2\gamma^2u_0^2\rb}{m^2\gamma^3}
& 
k\ll k_*.
\end{cases}
\ee
The sign change predicts instability threshold $u=u_c$ 
with the unstable and stable phases for $\left| p\frac{d\gamma}{dn}\right|-\gamma m u_0$ 
positive and negative, respectively. The sign change occurs simultaneously for $k\gg k_*$ and $k\ll k_*$. 
This gives critical velocity $u_c$ given in Eq.\eqref{eq:u_critical}.
If $R\sim 1$, 
the critical velocity is comparable to the plasmonic velocity $u_0$ in a dissipation-free system at density close to CNP. 


An estimate of the most unstable mode $k_\star$ follows from matching the small- and large-$k$ behavior 
Eq.\eqref{eq:zeta_large-k}: 
%
\be
\frac{k_*^2}{m^2\gamma^3}\lb p^2 \lp \frac{d\gamma}{dn}\rp^2 
-m^2\gamma^2u_0^2\rb=
\frac{\left| p\frac{d\gamma}{dn}\right| }{2m u_0}-\frac12\gamma
.
\ee
Solving for $k_*$ and noting that both 
sides vanish at the transition but remain nonzero away from transition, in the limit 
$\left| p\frac{d\gamma}{dn}\right| \to mu_0 \gamma$ we obtain \cite{footnote_Dn_Dp}: 
\be
\label{eq:k*_result}
k_*^2\approx \frac14\frac{\gamma^2}{u_0^2}
.
\ee
Notably, the maximally unstable wavelength $\lambda_\ast = 2\pi/k_\ast$ remains finite at the transition, in agreement with our simulations and in analogy with the finite modulation wavelength at the onset of the hydrodynamic Kapitsa instability.\cite{Andreev1964,Balmforth2004} Density-dependent $\gamma$ and $u_0$ yield a tunable critical velocity $u_c$ and wavelength ranging from submicron to hundreds of microns. The resulting narrow-band radiation frequency, $f = u_d/\lambda$, can thus span from sub-GHz to hundreds of GHz for realistic parameters; see \cite{Supplemental Information} for estimates and plots of the critical velocity and modulation wavelength.

In summary, we identify an electron–hydrodynamic instability in Dirac materials closely analogous to the Turing–Kapitsa instability that generates roll-wave trains in viscous films. Density dependence of current dissipation near charge neutrality makes the flow unstable above a critical drift velocity $u_c$, 
resulting in self-sustained running waves. This leads to a sharp, second-order–like change in the time-averaged current and narrow-band emission at the washboard frequency $f = u/\lambda$, with a current-tunable range extending to hundreds of GHz, highlighting Dirac materials as a promising platform for high-frequency electron-fluid phenomena.

This work greatly benefited from discussions with Ray Ashoori, Serguei Brazovskii, Andre Geim, Frank Koppens, Leo Radzihovsky, Brian Skinner, and Eli Zeldov.

\section{Supplemental Information}
\subsection{Numerical procedure} 

In this section, we outline the numerical procedure used to solve the hydrodynamic transport equations governing the primary variables: the particle density \( n(x,t) \) and the momentum density \( p(x,t) = m n u \). These equations can be written in the following form:
\begin{align}
\partial_t n &= -\partial_x \frac{p}{m} + D_n \partial_{x}^2 n, \label{eq:n_eq} \\ 
\partial_t p &= -\gamma(n) p 
- \partial_x \left( \frac{U}{2}  n^2 + \frac{p^2}{m n} \right) 
+ e n E 
+ D_p \partial_{x}^2 p, 
\nonumber
\end{align}
where 
$E$ is a spatially uniform driving field, $U$ is electron-electron interaction strength, $D_p$ and $D_n$ are diffusivity parameters introduced to regularize the dynamics (and taken to be small and positive). 
In simulations, we non-dimensionalize the problem by setting $m=e=1$ and, in this paper, work with $n(x)$ and $p(x)$ defined on a finite interval $0<x<L$, obeying Eqs.\eqref{eq:n_eq} 
with periodic boundary conditions, $n(0,t)=n(L,t)$ and $p(0,t)=p(L,t)$ for all $t$.

The results shown in Figs. 2 and 3 of the main text have been produced by the numerical procedure summarized below.
We discretize space, and integrate in time adaptively.
Spatial derivatives are computed in Fourier space (e.g., \cite{trefethen2000,boyd2001,canuto1988}) for high accuracy on periodic domains, and time integration uses a stiff Backward Differentiation Formula (BDF) method~\cite{hairerwannerII}. The solver automatically adjusts the time step to satisfy the prescribed error tolerances. 
Throughout, we enforce a density floor $n_{\mathrm{eff}}(x,t)=\max\{n(x,t),n_{\mathrm{floor}}\}$ with $n_{\mathrm{floor}}=10^{-7}$.
All terms requiring division by $n$ (e.g., $p/(mn)$ and $p^2/(mn)$) and all evaluations of $\gamma(n)$ and the pressure use $n_{\mathrm{eff}}$.
This avoids numerical singularities and has no discernible effect on converged solutions.

The functions $n(x)$ and $p(x)$ were represented as discrete Fourier series evaluated on an equispaced grid of 
$N_x$ points:
\be
\{ x_j=j\,L/N_x, \quad  j=0,\dots,N_x-1, \quad  \Delta x= L/N_x \}.
\ee
The number of operations required for each evaluation of the time derivative vector $(\p_t p,\p_t n)$
scales as $\mathcal{O}(N_x \log N_x)$ due to FFTs. Because accurate wave resolution is required, we choose $N_x$ so that the smallest dynamically relevant wavelength is resolved by at least $\sim$30--50 points; for the runs shown in Figs.~2--3 we used $L=10$ and $N_x=1024$.

For improved numerical accuracy, we compute the spatial derivatives in Eqs.~\eqref{eq:n_eq} 
in Fourier space. For a given target accuracy, this allows us to use fewer grid points than low-order finite differences, which can reduce the overall computational cost considerably. 
Specifically, let $\hat f_n$ denote the discrete Fourier transform of $f(x)$ on the grid
$x_j=j\Delta x$ ($j=0,\dots,N_x-1$), with $\Delta x=L/N_x$ and Fourier modes
\begin{equation}
k_n=\frac{2\pi}{L}
\begin{cases}
n, & n=0,1,\dots,\lfloor N_x/2\rfloor,\\
n-N_x, & n=\lfloor N_x/2\rfloor+1,\dots,N_x-1 .
\end{cases}
\end{equation}
The spatial derivatives are computed by multiplying Fourier coefficients and applying the inverse discrete Fourier transform:
\begin{align}
\partial_x f &= \Re\!\left\{\mathcal{F}^{-1}\!\left[\, i k_n\, \hat f_n \right]\right\},\\
\partial_x^2 f &= \Re\!\left\{\mathcal{F}^{-1}\!\left[\, -k_n^2\, \hat f_n \right]\right\}.
\end{align}

Nonlinear products (such as the pressure and damping contributions) are formed pointwise in $x$ space and returned to $k$ space by Fast Fourier Transform (FFT).

To control aliasing from these nonlinearities we apply the Orszag 2/3 truncation rule \cite{orszag1971}.
Specifically, whenever $\partial_t n$ and $\partial_t p$ are assembled on the grid, we transform them to Fourier space, set all modes with $|m|>N_x/3$ to zero (equivalently $|k|>(2/3)k_{\max}$ with $k_{\max}=\pi/\Delta x$), and transform back before passing the derivatives to the time integrator.

On a finite grid, quadratic products generate Fourier interactions beyond the Nyquist wavenumber \cite{canuto1988}. These spurious components appear at incorrect lower frequencies or wavelengths unless they are filtered out.
Keeping only the lowest two–thirds of modes guarantees \cite{orszag1971} that every quadratic interaction of retained modes either stays within the allowed range or falls completely outside it, where it is removed.

We use the method of lines (MOL), discretizing the spatial coordinate to convert the PDEs into a system of ordinary differential equations (ODEs) in time.
The system is integrated in time using a stiff \emph{BDF} method \cite{hairerwannerII} as implemented in the SciPy package \cite{virtanen2020scipy}. The solver uses adaptive internal time stepping, while we record solution snapshots at equally spaced output times
$t_j = j\,t_{\mathrm{final}}/(n_{\mathrm{save}}-1)$ for $j=0,1,\dots,n_{\mathrm{save}}-1$. 
The BDF method automatically adapts both the time step and the method order to satisfy the prescribed error tolerances.

Parameter values used in the simulations to obtain the results shown in Figs.~2 and 3 were
$U=1$, $\bar n=0.2$, $\gamma_0=2.5$, $w=0.04$,
$u_d=5.245$, $D_n=0.5$, $D_p=0.1$,
$L=10$. Parameters used in the FFT procedure described above were $N_x=1024$, $t_{\mathrm{final}}=50$, $n_{\mathrm{save}}=100$,
with the density floor $n_{\mathrm{floor}}=10^{-7}$; we used relative and absolute tolerances $\mathrm{rtol}=10^{-4}$ and $\mathrm{atol}=10^{-7}$. In addition, to control aliasing, we used the Orszag $2/3$ truncation rule. Specifically, each time the time-derivative fields $\partial_t n$ and $\partial_t p$ are assembled on the grid, we discard the highest spatial Fourier harmonics (the upper third of the spectrum) before passing these derivatives to the time integrator.
These tolerance values were determined through numerical experiments as optimal: reducing them further increases runtime, while loosening them may introduce numerical artifacts. Therefore, when accuracy concerns arise, 
we refine the spatial discretization $N_x$ and adjust the tolerance parameters accordingly.

The BDF integrator automatically adjusts the time step (and the method order) using a local error estimate so that the weighted local truncation error remains within the specified tolerances. The spatial discretization is fixed once a run begins: both the number of grid points $N_x$ and the domain length $L$ remain constant for the duration of the simulation. 

To optimize the run time for achieving instability, the domain length $L$ was chosen so that the set of discrete harmonics 
\be
\{k_m=2\pi m/L, \quad m=1,2...\}
\label{eq:k_set}
\ee 
contains the most unstable harmonic $k_*$ (see Fig. 1 of the main text). Once chosen, $L$ and $N_x$ were kept fixed during the simulation.
We first pick an integer 
$1<M<15$ representing the intended number of wavelengths in the box and set
\begin{equation} 
 L=\frac{2\pi M}{k_*}
\end{equation}
so that $k_*$ coincides exactly with the harmonic $k_M$ in the set defined above, Eq.\eqref{eq:k_set}.
We then select $N_x$ large enough such that the set of harmonics given in Eq.\eqref{eq:k_set} includes not only $k_*$ but also a few of the lower harmonics; 
in practice we enforce $M\le N_x/9$, which ensures that after 2/3 de-aliasing (retaining $|m|\le N_x/3$) the first three harmonics of the dominant mode remain resolved: $3M \le N_x/3$.

The box size and initial conditions are fixed before the run and are not altered during the simulation.

Simulations start from a spatially uniform drifting state, $n(x,0)=\bar n$ and $p(x,0)=m\bar nu_d$.
To seed pattern formation we add a small, deterministic superposition of sine and cosine modes, 
\begin{align}
n(x)_{t=0}&=\bar n+\sum_{j}(a_j\cos k_j x+b_j\sin k_j x),\\ 
p(x)_{t=0}&=m\bar n u_d+\sum_{j}(a'_j\cos k_j x+b'_j\sin k_j x).
\end{align}
where $k_j=\frac{2\pi j}L$ 
with $j$ spanning a finite set of low integers distributed uniformly on a log scale (for example $\{3,5,8,13,21\}$) and moderate values of the coefficients $a_j$, $a'_j$, etc., were chosen, typically $a_j, b_j \lesssim0.2\bar{n}$ and $a'_j, b'_j \lesssim0.2m\bar{n}u_d$. We verified that the late-time patterns are insensitive to the precise seed choice within this small-perturbation regime.

The density-dependent damping $\gamma(n)$ and the pressure term are taken as defined in the main text, with $n$ evaluated as $n_{\mathrm{eff}}=\max(n,n_{\mathrm{floor}})$ in these expressions.
The driving field $E$ is uniform in space; for constant-field runs we set $E=(m u_d/e)\langle \gamma(\bar n)\rangle$.

\subsection*{Current Relaxation due to momentum-conserving Scattering in a Two–Band Semimetal}
\label{sec:current_relaxation}

Electron–electron (ee) interactions conserve total momentum but need not
conserve electric current.  In a Galilean-invariant system,
$\vec v_{\vec k}=\vec p_{\vec k}/m$, current is proportional to
momentum and hence cannot be relaxed by momentum-conserving collisions.
In multiband semimetals such as bilayer graphene (BLG), however, Galilean
invariance is broken: band velocities are not proportional to crystal
momentum, and electrons and holes may contribute oppositely to the current
even at the same momentum. As a result, ee scattering can relax current
without relaxing total momentum. Below we separate and compare the distinct
mechanisms that contribute to current relaxation and give compact criteria
for their relative importance.

\subsection*{Two-band model and interband (electron–hole) current relaxation}

We begin with the low-energy two-band description of BLG,
\begin{align}
\varepsilon_{\pm}(\vec k)
    &= \pm \tfrac12\!\left(\sqrt{\gamma_1^2 + 4(\hbar v k)^2} - \gamma_1\right),
\label{eq:BLG_spectrum}\\
\vec v_{\pm}(\vec k)
    &= \nabla_{\vec k}\varepsilon_{\pm}(\vec k)
     = \pm \frac{2\hbar v^2\,\vec k}{\sqrt{\gamma_1^2+4(\hbar v k)^2}},
\label{eq:BLG_velocity}
\end{align}
with $v\approx 10^6\,$m/s and $\gamma_1=0.39\,$eV.  At charge neutrality
($\mu=0$) conduction and valence states at the same momentum satisfy
$\vec v_+(\vec k)=-\vec v_-(\vec k)$. Consequently a
momentum-conserving collision that transfers an excitation between the two
bands reverses its current contribution:
\[
\Delta\vec J
  = e\big[\vec v_{s_1'}(\vec k_1')+\vec v_{s_2'}(\vec k_2')
         -\vec v_{s_1}(\vec k_1)-\vec v_{s_2}(\vec k_2)\big]\neq0,
\]
even though $\vec k_1+\vec k_2=\vec k_1'+\vec k_2'$. Thus
interband (electron–hole) processes provide an efficient channel for
current relaxation near neutrality where thermally excited electrons and
holes coexist.  A simple estimate of the corresponding current relaxation rate in
the nondegenerate regime ($\mu\ll k_B T$) yields the Planckian scale
\[
\Gamma_J\sim \frac{k_B T}{\hbar},
\]
up to dimensionless phase-space and matrix-element factors.

\subsection*{Intraband relaxation away from neutrality: the $\eta$-model}

For $\mu\gg k_B T$ interband processes are exponentially suppressed, yet
momentum-conserving collisions can still relax current because
$\vec v_{\vec k}\neq\vec p_{\vec k}/m_*$. To quantify this
mechanism introduce the Fermi-surface mismatch
\[
\boldsymbol\chi(\theta)
   = \vec v(k_F,\theta) - \frac{\hbar\vec k_F(\theta)}{m_*},
\]
where $m_*$ is chosen so that the Fermi-surface averages match:
$\langle\vec v\rangle_{FS}=\langle\vec p/m_*\rangle_{FS}$. Define the
dimensionless non-parabolicity parameter
\begin{equation}
\eta^2 = \frac{\langle|\boldsymbol\chi(\theta)|^2\rangle_\theta}
                {\langle|\vec v(k_F,\theta)|^2\rangle_\theta}.
\label{eq:eta_def}
\end{equation}
Only the current component that projects onto the non-Galilean subspace
(proportional to $\boldsymbol\chi$) can decay under momentum-conserving
collisions, while the momentum-tied part is protected. Thus a minimal
phenomenological form for the intraband current-relaxation rate is
\begin{equation}
\Gamma_{\eta}\;\approx\;\gamma_{ee}(T,\mu)\,\eta^2(n),
\label{eq:Gamma_eta}
\end{equation}
where $\gamma_{ee}$ is the microscopic quasiparticle (ee) scattering rate.
Using the four-band BLG dispersion (Eqs.~\ref{eq:BLG_spectrum}--\ref{eq:BLG_velocity})
one finds that $\eta(n)$ grows with density (see below and Fig.~\ref{fig:eta_vs_n}):
in the window $n=10^{12}$–$5\times10^{12}\,$cm$^{-2}$ we obtain
$\eta\approx0.16$–$0.67$, so $\eta^2\sim 0.03$–$0.45$.

\subsection*{Analytic estimate for \(\eta\) and parabolic limit}
\begin{figure}[t]
\centering
\includegraphics[width=0.95\linewidth]{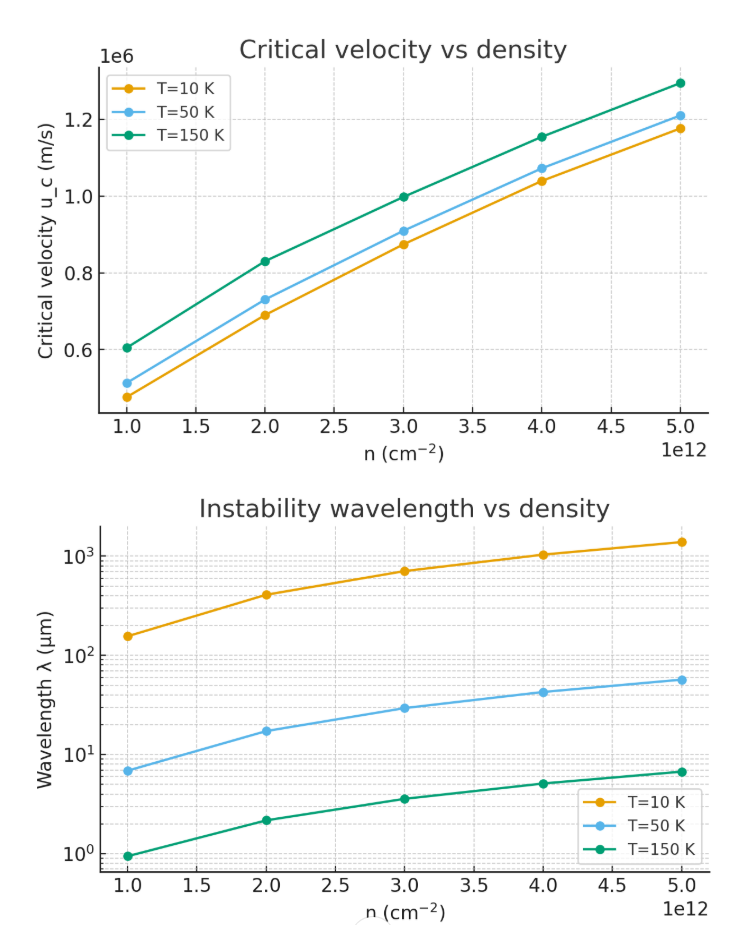}
\caption{Critical velocity and modulation wavelength vs. electron density at the Turing-Kapitsa instability for BLG band, Eq.\eqref{eq:BLG_spectrum}. The results shown were obtained using the current relaxation rate $\Gamma_J$ given in Eq.\eqref{eq:Gamma_full}. }
\label{fig:eta_vs_n}
\end{figure}

A compact analytic relation follows from the four-band kinematics:
evaluating radial velocities at the Fermi contour gives
\begin{align}
& v(k_F)=\frac{2\hbar v_F^2 k_F}{Z},\quad
\frac{p}{m_*}=\frac{\hbar k_F}{m_*}=\frac{2\hbar v_F^2 k_F}{\gamma_1},
\\
\quad 
&Z=\sqrt{\gamma_1^2+4(\hbar v_F k_F)^2},
\end{align}
so that
\[
\frac{|\chi|}{v} \simeq \frac{2\varepsilon_F}{\gamma_1},
\qquad\Rightarrow\qquad
\boxed{\;\eta\simeq \frac{2\varepsilon_F}{\gamma_1},\quad
\eta^2\simeq\left(\frac{2\varepsilon_F}{\gamma_1}\right)^2\;},
\]
with $\varepsilon_F$ the conduction-band Fermi energy measured from the
band edge. In the low-energy (parabolic) limit 
\[
\varepsilon_F\simeq
\pi\hbar^2 n/(g m_*)+k_BT/a\] 
(with $g=4$) this gives the linear-in-density
estimate
\[
\eta=\frac{2\varepsilon_F}{\gamma_1}
\]
but the exact $\eta\simeq2\varepsilon_F/\gamma_1$ (used in the plots)
should be used once $k_F$ approaches the crossover scale $\sim\gamma_1/\hbar v_F$.
Figure~\ref{fig:eta_vs_n} compares the exact $\eta(n)$, the parabolic
estimate, the cyclotron mass $m_c(n)$ and $\varepsilon_F(n)$.

\subsection*{Collinear (head-on) scattering: suppression of odd-harmonic relaxation}

Recent kinetic analyses of 2D Fermi liquids show that Pauli blocking and
2D kinematics bias surviving two-particle collisions toward nearly
collinear (head-on) events: incoming momenta are nearly opposite and
outgoing momenta are nearly opposite, with small angular deflections
$\delta\theta\sim\mathcal{O}(k_B T/\mu)$ when $k_B T\ll\mu$
\cite{Ledwith2017,Ledwith2019,Kryhin2021}. Because current is carried by
odd angular harmonics on the Fermi surface, relaxation of those harmonics
requires angular diffusion. Small per-collision angular steps imply that
odd-mode relaxation is parametrically slower than the quasiparticle
collision rate; to leading order the odd-mode relaxation rate carries a
suppression factor scaling as $(k_B T/\mu)^2$ in the degenerate limit.

A practical way to include this effect is to multiply the $\eta$-model
estimate~\eqref{eq:Gamma_eta} by a collinearity suppression factor
$S_{\rm col}(T,\mu)$:
\begin{equation}
\Gamma_J(T,\mu)\;\equiv\;\tau_J^{-1}(T,\mu)
    \;\approx\; \gamma_{ee}(T,\mu)\;\eta^2(n)\;S_{\rm col}(T,\mu).
\label{eq:Gamma_full}
\end{equation}
A minimal interpolant that captures the asymptotics is
\begin{equation}
S_{\rm col}(T,\mu)\;=\;
C\lp \frac{k_BT}{\mu +\frac{k_BT}{a}}\rp^2
\qquad C\sim\mathcal O(1),
\label{eq:Scol}
\end{equation}
so $S_{\rm col}\to1$ for $k_B T\gtrsim\mu$ (nondegenerate regime) and
$S_{\rm col}\propto (k_BT/\mu)^2$ for $k_BT\ll\mu$.  The constant $C$
can be set to unity for order-of-magnitude estimates or determined by a
microscopic collision-integral calculation.

\subsection*{Interpolating quasiparticle rate and combined scaling behavior}

To have a single expression valid both in degenerate and nondegenerate
regimes we use the interpolant
\begin{equation}
\gamma_{ee}(T,\mu)
  = \frac{k_B T}{\hbar}\;\frac{k_B T}{k_B T + a\,\mu},\qquad a\simeq1,
\label{eq:gamma_interp}
\end{equation}
which yields $\gamma_{ee}\simeq k_BT/\hbar$ for $\mu\ll k_BT$ and
$\gamma_{ee}\simeq (k_BT)^2/(\hbar\mu)$ for $k_BT\ll\mu$.

Combining Eqs.~\eqref{eq:Gamma_full} and \eqref{eq:gamma_interp} gives
the two limiting behaviours:

\medskip\noindent
\textbf{(i) Nondegenerate / near neutrality ($\mu\ll k_B T$):}
\[
\Gamma_J\sim \frac{k_B T}{\hbar}\,\eta^2(n)\,S_{\rm col}\simeq \frac{k_B T}{\hbar}\,\eta^2(n),
\]
since $S_{\rm col}\to1$.  Interband electron–hole processes also give a
rate $\Gamma_{\rm inter}\sim k_B T/\hbar$; near neutrality both channels
are active and comparable up to phase-space/matrix-element factors.

\medskip\noindent
\textbf{(ii) Degenerate ($k_B T\ll\mu\approx\varepsilon_F$):}
\begin{align}
& \Gamma_J \;\approx\; \frac{(k_B T)^2}{\hbar\mu}\,\eta^2(n)\;S_{\rm col}
\;\simeq\;
\frac{(k_B T)^2}{\hbar\mu}\,\eta^2(n)\;\times\;
\left(\frac{k_B T}{\mu}\right)^2
\nonumber \\
& \;=\;\frac{(k_B T)^4}{\hbar\,\mu^3}\,\eta^2(n),
\end{align}
using $S_{\rm col}\propto (k_B T/\mu)^2$ in the degenerate limit. Using
the estimate $\eta^2\propto\varepsilon_F^2\sim\mu^2$ (exactly,
$\eta\simeq2\varepsilon_F/\gamma_1$), the combined $\mu$-dependence is
\[
\Gamma_J\propto\frac{\mu^2}{\mu^3}=\mu^{-1}.
\]
Thus \emph{with} collinear suppression included the intraband current
relaxation rate decreases with increasing $\mu$ (and density) in the
degenerate regime.  By contrast, \emph{without} collinearity suppression
($S_{\rm col}=1$) one would find $\Gamma_{\eta}\propto\mu$ (i.e. increasing
with density) because $\gamma_{ee}\propto (k_BT)^2/\mu$ while $\eta^2\propto\mu^2$.

\subsection*{Which mechanism dominates? — a short decision guide}

\begin{itemize}
\item \emph{Near neutrality} ($\mu\lesssim k_B T$): interband electron–hole
      scattering dominates current relaxation; expect \(\Gamma_J\sim k_B T/\hbar\).
\item \emph{Moderate doping} (intermediate $\mu$ where $\eta$ becomes
      appreciable but $k_B T$ is not extremely small): intraband non-parabolic
      processes (the $\eta$-mechanism) can contribute significantly and may
      even compete with interband rates if $S_{\rm col}\approx1$.
\item \emph{Deeply degenerate regime} ($k_B T\ll\mu$): collinear/head-on
      kinematics strongly suppress odd-harmonic relaxation; the net
      $\Gamma_J$ typically decreases with density. 
\end{itemize}

A practical criterion to check in any parameter set is
\[
\text{interband dominates if}\quad
\frac{k_B T}{\hbar}\;>\;\gamma_{ee}\,\eta^2\,S_{\rm col},
\]
with $\gamma_{ee}$ given by Eq.~\eqref{eq:gamma_interp} and $S_{\rm col}$
by Eq.~\eqref{eq:Scol}.

\subsection*{Application to the TK instability and numerical workflow}

In the TK analysis we use the final, collinearity-corrected rate
\(\Gamma_J(T,n)\) from Eq.~\eqref{eq:Gamma_full}. The relevant TK quantities
are computed exactly as in the main text:
\[
u_0(n)=\sqrt{\frac{U n}{m^*}},\qquad
R=\frac{d\ln\gamma}{d\ln n},\qquad
u_c=\frac{u_0}{|R|},
\]
and
\[
k_*=\frac{\gamma}{2u_0},\qquad \lambda=\frac{2\pi}{k_*},
\]
with $\gamma\equiv\Gamma_J$ used for the instability formulas. 

\subsection*{Interaction strength}

The Thomas–Fermi (long-wavelength) interaction scale entering
$u_0(n)=\sqrt{U n/m^*}$ is the inverse compressibility of a 2D parabolic
band,
\[
U=\frac{2\pi\hbar^2}{g\,m^*},
\]
with $g$ the total degeneracy (spin×valley). For BLG with
$m^*=\gamma_1/(2v_F^2)$ and $g=4$ this yields
$U=\pi\hbar^2/(2m^*) = 2\pi\hbar^2 v_F^2/\gamma_1$.

\subsection*{Concluding remarks}

The decomposition above shows three distinct, physically transparent
contributions to current relaxation: (i) interband electron–hole processes
dominant near neutrality, (ii) intraband non-parabolicity (the $\eta$-effect)
which can be important at moderate $\mu$, and (iii) strong collinear
suppression of odd-harmonic relaxation in the degenerate limit.  Which
mechanism controls transport depends sensitively on $k_B T/\mu$ and on the
density through $\eta(n)$; the expressions and decision criteria given here
permit a straightforward and reproducible evaluation across parameter space.


\begin{thebibliography}{9}
\bibitem{footnote_Pi_Phi} This is achieved using the identity 
\[ 
n(x)\p_x\Phi(x)=n(x)\p_x  U_0(n(x)-\bar n) =\p_x\frac{U_0}2 n^2(x)
\] 
where density offset parameter $\bar n$ originates from Eq. \eqref{eq:Phi(x)}  
where it was introduced to maintain charge neutrality of the unperturbed equilibrium state.
As a quick check on above result and Eq. \eqref{eq:Phi(x)}, 
take a system with no electric field or current, perturbed by a static, position-dependent background $\bar n(x)$. 
The above relation predicts the response of the form
$n(x)=\frac{U_0}{U}\bar n(x) $, in agreement with the Thomas–Fermi screening picture. 
As expected, for a system at constant density, $\bar n$ drops out from equations of motion, reflecting that only the electrons are moving under electric field, whereas the background charges contributing to $\bar n$ remain stationary.
\bibitem{footnote_Dn_Dp}
Finite $D_n$ and $D_p$ may shift the instability threshold and alter $\lambda_*$. However, at small $D_n$ and $D_p$ this effect is expected to be small.
\end{thebibliography}
\end{document}